\documentclass{article}

\usepackage{arxiv}
\usepackage{amsmath}
\usepackage{algorithm}
\usepackage{algpseudocode}
\usepackage{appendix}
\usepackage[utf8]{inputenc} 
\usepackage[T1]{fontenc}    
\usepackage[colorlinks=true,linkcolor=red,urlcolor=blue,citecolor=blue]{hyperref}       

\usepackage{url}            
\usepackage{booktabs}       
\usepackage{amsfonts}       
\usepackage{cuted} 
\usepackage{nicefrac}       
\usepackage{microtype}      
\usepackage{lipsum}
\usepackage{graphicx}
\graphicspath{ {./images/} }
\usepackage{setspace}
\usepackage{epstopdf}

\title{multi-scale traffic flow modeling: a renormalization group theoretic approach}

\author{
 Zhaohui Yang \\
  \texttt{zhaohuiyang@google.com} \\ 
   \and
 Kshitij Jerath \\
  \texttt{Kshitij\_Jerath@uml.edu} \\
}

\begin{document}

\maketitle
\begin{abstract}

Traffic flow modeling is typically performed at one of three different scales (microscopic, mesoscopic, or macroscopic), each with distinct modeling approaches. Recent works that attempt to merge models at different scales have yielded some success, but there still exists a need for a single modeling framework that can seamlessly model traffic flow across several spatiotemporal scales. The presented work utilizes a renormalization group (RG) theoretic approach, building upon our prior research on statistical mechanics-inspired traffic flow modeling. Specifically, we use an Ising model-inspired cellular automata model to represent traffic flow dynamics. RG transformations are applied to this model to obtain coarse-grained parameters (interaction and field coefficients) to simulate traffic at coarser spatiotemporal scales and different vehicular densities. We measure the accuracy of the coarse-grained traffic flow simulation using a pixel-based image correlation metric and find good correlation between the dynamics at different scales. Importantly, emergent traffic dynamics such as backward moving congestion waves are retained at coarser scales with this approach. The presented work has the potential to spur the development of a unified traffic flow modeling framework for transportation analysis across varied spatiotemporal scales, while retaining an analytical relationship between the model parameters at these scales.
\end{abstract}



\section{\label{sec:level1}Introduction}

Modeling, predicting, and controlling traffic flow is a necessity in the modern world, since the economic and productivity costs of congestion continue to rise. In addition, reduction in congestion is accompanied with several benefits, such as fewer traffic incidents and reductions in emissions.
Consequently, one of the primary goals in the study of modern transportation systems is to devise strategies that minimize congestion and ensure smooth flow of traffic \cite{papageorgiou2003review}. 
To examine and influence vehicular traffic flow, the field has typically relied on a few primary forms of modeling, either based on first principles or using data-driven approaches \cite{Williams1998urban, Lv2015traffic, rodriguez2005first}. Traditionally, first principles-based traffic flow models study the system at three different spatial scales: microscopic, macroscopic, or mesoscopic. 
    
Microscopic-scale car-following models describe how a driver (or algorithm) controls the longitudinal vehicle dynamics (i.e. speed and position) relative to a preceding vehicle on a highway. These models use ordinary differential equations (e.g., Intelligent Driver Model (IDM) \cite{Kesting2010enhanced}\cite{zhou2016impact})
or cellular automata (CA) for microscopic-scale traffic flow modeling
\cite{nagel1992cellular}\cite{barlovic1998metastable}\cite{Jerath_conferenceACC_2014}. 
These types of models help encapsulate significant detail, 
not all of which may be relevant to model, predict, or control emergent patterns in traffic flow. Moreover, large-scale microscopic-scale modeling and simulations are often accompanied with significant computational costs, and may potentially lack analytical insight (due to the reliance on numerical simulations).
On the other hand, macroscopic-scale models treat traffic flow as a continuum and use partial differential equations to focus on the dynamics of vehicular density $\rho(x,t)$ and/or flow $q(x,t)$ as functions of space $x$ and time $t$. First developed by Lighthill and Whitham, these models compared the traffic flow to `flood movement' \cite{lighthill1955kinematic}. 
Though the macroscopic-scale dynamical properties are reliably represented in these models, the potential impacts of specific microscopic-scale actions by individual vehicles are sometimes lost. As a result, these models are generally not well-suited for modeling heterogeneous traffic flows, though there have been some recent works to address this limitation \cite{delis2016simulation}\cite{yang2022macroscopic}\cite{yang2020observability}.

In contrast to microscopic- and macroscopic-scale modeling approaches, mesoscopic-scale models have been proposed as a means to capture dynamics associated with intermediate scales \cite{mahnke2005probabilistic, jerath2010adaptive, yang2018examining, Kimconference2016}. The mesoscopic modeling approach was first proposed by Ilya Prigogine using evolution of speed and position probability distributions \cite{prigogine1960boltzmann}. Recently, these approaches have expanded to include modeling traffic flows as queues, studying the dynamics of road links (such as in the cell transmission model \cite{Daganzo1994cell}), or examining the behavior of vehicular clusters using probabilistic master equations \cite{mahnke2005probabilistic}\cite{jerath2015dynamic}. 

While each of these approaches model several aspects of the traffic flow dynamics very well at their respective spatial scales, they leave some fundamental questions unanswered: how does the choice of spatial scale influence the ability to effectively and accurately model emergent behaviors, such as self-organized congestion \cite{nagel1992cellular}, in traffic flow? Are specific spatial scales better suited for modeling, quantifying, and predicting traffic flow dynamics? 
This paper seeks to address some of these questions using renormalization group theory. 

\subsection{Recent Works}
    
\label{SubSec:Recent_works}
        Some recent works have sought to generate a link between the microscopic-level description and macroscopic-scale dynamics of traffic flow, with varying degrees of success \cite{cristiani2012can}\cite{teoh2018renormalization}. For example, the Cell Transmission Model (CTM) has been used to model short sections of the roadways (called cells), where the size of a single cell is equal to the distance traveled by a vehicle in light traffic in a single tick (i.e. a single simulation time step) \cite{Daganzo1994cell}. This representation has been successfully used to predict temporal evolution of traffic, including transient phenomena such as the building, propagation, and dissipation of queues at a fixed spatial scale. While there have been attempts to modify this approach to account for cells of varying sizes (which could potentially allow for selection of an `appropriate' spatial scale for modeling and prediction), these have been met with considerably less interest than other applications of CTM \cite{ziliaskopoulos1997cell}. 
        
        
        Other recent works, such as variable multi-scale modeling by Cristiani et al. \cite{cristiani2012can}, have attempted to describe traffic dynamics using both a discrete (microscopic-scale) and a continuous (macroscopic-scale) perspective simultaneously, through the coupling of models at different spatial scales. Such a weighted coupling methodology allows prioritization of models across a continuous parameter space and could potentially enable modeling at an arbitrary choice of spatial scales, but still relies on two different modeling frameworks. 
        Han and Ee have sought to use the Renormalization Group (RG) to create such a relationship between spatial scales, but their work relies on numerical approaches to perform renormalization on the Nagel-Schreckenberg model \cite{teoh2018renormalization}. Thus, there exists a need to provide a single systematic framework to model traffic flow across multiple spatial scales, for which closed-form recursive  relationships between model parameters at different scales can be determined.
        

        In subsequent sections we build on our prior work on statistical mechanics-inspired traffic flow modeling \cite{jerath2015dynamic, Yang2020, jerath2010impact}, and leverage Renormalization Group (RG) theory to model system dynamics at several different spatial scales.
        Specifically, in Section \ref{Sec:2-Statisitcal-mechanics} 
        we demonstrate how our statistical physics-inspired approach can be used to model emergent traffic flow dynamics.
        Section \ref{Sec:3-Traffic-flow} describes how Renormalization Group theory can be applied to this model to simulate traffic at coarser spatial scales, as well as generate a link between the microscopic and macroscopic scales. We demonstrate the mechanism to relate the interaction and field coefficients used to model traffic flow dynamics at different spatial scales \cite{wilson1975renormalization}. In Section \ref{Sec:Results}, we show numerical examples to validate the Renormalization Group theoretic approach to modeling traffic dynamics at different scales, as well as quantify its accuracy. Section \ref{Sec:4-Concluding-remarks} summarizes the results and describes the future potential of this approach to benefit control and estimation algorithms in transportation systems. 
        

\section{Statistical Mechanics-inspired Cellular Automata Traffic Model}

\label{Sec:2-Statisitcal-mechanics}
Statistical physics can be used to understand large-scale systems using statistical descriptions as well as describe phase transitions and behaviors near criticality \cite{pathria1996statistical}. This makes its underlying concepts uniquely suited for application to emergent behaviors observed in traffic flow dynamics, including the occurrence of emergent patterns such as `phantom' traffic jams  \cite{nagel1992cellular}\cite{kerner1997experimental}. The statistical mechanics-based model presented in this work shares some features with Cellular Automata (CA) models, as it relies on a discrete-space, discrete-time representation of traffic flow on a closed ring road \cite{Nagel1992}.
Specifically, CA models allow us to capture microscopic-scale dynamics at a chosen spatial granularity (typically one car-length), but cells of alternative granular size can be used to simulate traffic flow dynamics at coarser spatial scales \cite{Daganzo1994cell}\cite{Hauck2014cellular}. 
In this paper, we leverage our prior work on statistical mechanics-inspired modeling  to systematically generate coarse-grained traffic flow models at different spatial scales using renormalization group theory \cite{jerath2015dynamic}\cite{Yang2020}. 



\begin{figure}[t]
\centering
  \includegraphics[width=0.6\columnwidth]{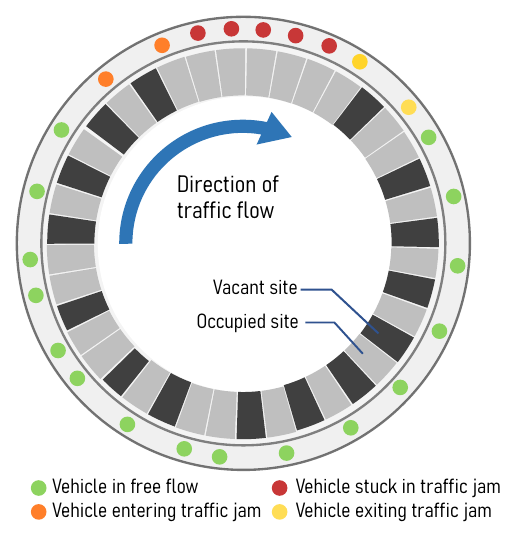}
	\caption{Cellular Automata (CA) traffic model, with occupied sites representing vehicles and vacant sites representing free space. Vehicle transitions from one site to another are implemented through spin exchange dynamics, and are inspired by statistical mechanics. Each site is the length of a single vehicle (approximately 5 m).
	}
	\label{fig:CA}
\end{figure}

\subsection{Statistical Mechanics-inspired Traffic Flow Dynamics}
A schematic representation of CA-based traffic flow model on a closed ring-road is shown in Fig. \ref{fig:CA}. Traditionally, CA models establish basic `rules' to imitate traffic system dynamics within this discrete-time, discrete-space environment \cite{Nagel1992}. We provide an alternative methodology to model the traffic dynamics that are inspired by statistical mechanics approaches, such as the Ising model \cite{glauber1963time}.

Similar to the one-dimensional Ising model, these dynamics can be modeled using $N$ sites, each spanning the space occupied by a single vehicle, and with states $S_i \in \{-1, +1\}$. The state $S_i=-1$ denotes a vacant site, and $S_i = +1$ denotes a site occupied by a vehicle, corresponding to the dark and light sites in Fig. \ref{fig:CA}, respectively. Further, let $\textbf{S} = \{S_1, S_2, ..., S_N\}$ denote the microstate of the system corresponding to various configurations of how vehicles are located at various sites on the ring-road. It is instructive to note that, for a given vehicular density $\rho = r/N$ where $r$ identical vehicles may be individually located on $N$ sites of the ring road, the cardinality of this set is given by $|\textbf{S}| ={N \choose r}$. It is reasonable to assume that no two vehicles occupy the same site at the same time. Further, at the finest scale, the length of each site is assumed to be length of a representative vehicle (typically around 5 m).
The Hamiltonian $H$ for the entire system is given by:
\begin{align} 
    H = -J\sum_{\langle i,j\rangle} S_iS_j-F\sum_{i=1}^N S_i
\label{Eq:Hamiltonian_Global}
\end{align}
\noindent where $S_i$ and $S_j$ represent the states of sites $i$ and $j$, and $\langle i,j\rangle$ serves as a pair of neighboring sites with index $i$ and $j$.
As in the Ising model, the Hamiltonian consists of two terms: the first term relates to interaction energy between a site and its neighbors quantified by the coefficient $J$, while the second term pertains to the field, whose strength is determined by the coefficient $F$.
In our work, the interaction and field terms are representative of interactions between neighboring vehicles and the average speed of traffic flow, respectively \cite{jerath2015dynamic}.
While the Hamiltonian expressed in Eq. (\ref{Eq:Hamiltonian_Global}) represents the energy of the entire system, typical vehicle driving behavior and dynamics are dominated by local effects. Thus, a site-based local Hamiltonian is better suited to model the transition probabilities associated with longitudinal vehicle movement along the roadway, and is given by:
\begin{align} \label{Hamiltonian_local}
    H_s(i) = -F S_i-J\sum_{j \in \mathcal{N}(i)} S_iS_j 
\end{align}
\noindent where $\mathcal{N}(i) = \{\:j: d(i, j)<d, \mbox{for }j<i\}$ represents the neighborhood of agent $i$, $d(i,j) = |(i-j)d_s|$ indicates the distance between sites $i$ and $j$, $d_s$ represents the size of a single site, and $d$ is the ‘look-ahead’ distance downstream of the location of site $i$. Since only the nearest neighbor has been considered here, site-based local Hamiltonian can be rewritten as:

\begin{equation}
    H_s(i) = -F S_i-J S_i S_{i+1} 
\end{equation}

The longitudinal movement of a vehicle along the roadway is modeled via spin dynamics as outlined in \textbf{Algorithm 1} \cite{jerath2015dynamic}. The literature outlines various potential spin dynamics to be used across a wide array of use cases. One choice could be spin-flip dynamics, wherein an occupied site with state $S_i = +1$ at time instant $k$ could `flip' to become a vacant site with state $S_i = -1$ at the next time instant $k+1$. However, this choice would cause the total number of vehicles on a closed ring-road to vary over time, an outcome that is impossible on a closed ring-road with no vehicles entering or leaving the system, i.e. where the total number of vehicles is conserved. Instead, we model vehicle motion using spin exchange dynamics, where adjacent sites can exchange their spin states, such that a vehicle moves forward by transforming a state pair from $\{S_i,S_{i+1}\} = \{+1, -1\}$ at time step $k$, to state pair $\{S_i,S_{i+1}\} = \{-1, +1\}$ at time step $k+1$.
Further, assuming that only one pair of spin states can be exchanged at each update or time step, the transition probability $\phi$ of a spin exchange occurring, i.e. a vehicle at site $i$ moving forward by one site, is given by:
\begin{equation}
\begin{aligned}
\label{Eq:Transition probability }
    \phi &= \: \mbox{exp}\left\{- \beta H_s(i)\right\}\\
\end{aligned}
\end{equation}
where $\beta$ is proportional to vehicular density, and the likelihood of spin exchange occurring is dependent on the local Hamiltonian $H_s(i)$. The spin exchange dynamics are implemented by comparing the transition probability to a randomly selected number from the uniform distribution $U(0,1)$. The effectiveness of the statistical mechanics-inspired CA traffic model has previously been evaluated using Monte Carlo simulations \cite{jerath2015dynamic}.

\begin{algorithm}[H]
\caption{Simulation of traffic flow using  modified Metropolis Algorithm}\label{euclid}
\begin{algorithmic}[1]
\State{Initial condition: \:}{Vehicles are randomly distributed along an N-site ring road with a given vehicle density $\rho$}
\While{$t< \mbox{\texttt{endTime}} $}
    \For{site number $i=1$ to $N$}
        \If {$S_i=+1$ and $S_{i+1}=-1$}
             \State{produce a random number $p \sim U(0,1)$}
             \State{calculate local Hamiltonian $H_s$ of site $i$ }
             \State{calculate transition probability $\phi=e^{-\beta H_s }$}
             \If {$p<\phi$}
                \State{exchange spin states $S_i$ and $S_{i+1}$}
          \EndIf
        \EndIf
    \EndFor
\State $t \gets t+1$
\EndWhile\label{euclidendwhile}
\end{algorithmic}
\end{algorithm}

\section{Traffic flow modeling across multiple scales}
\label{Sec:3-Traffic-flow}

In this section, we apply concepts from renormalization group (RG) theory to the statistical mechanics-inspired traffic model and generate coarse-scale models of traffic flow dynamics at several different spatial scales. A key component of the RG approach is to build the  coarse-grained description using the same functional form of the Hamiltonian as expressed in Eq. \ref{Eq:Hamiltonian_Global}, but with modified coefficients that can be related to the coarse-grained sites and states.

The system Hamiltonian (Eq. \ref{Eq:Hamiltonian_Global}) provides insights into how the interactions between neighboring sites influence the energy of the entire system. Macroscopic properties other than energy can be analyzed using the partition function. The partition function describes the statistical properties of a system in thermodynamic equilibrium. For the set $\mathbf{S} =\{S_1, S_2, ..., S_N \}$ of system microstates, the partition function for the Ising model-based ring-road traffic system is given by:
\begin{align} 
\begin{aligned}
    Z = \sum_{\mathbf{S}}e^{-H/kT} 
\label{Eq:Partition function 1}
\end{aligned}
\end{align}

\noindent where $H$ denotes the system Hamiltonian, $k$ denotes Boltzmann's constant, and $T$ is proportionally determined by the traffic density. During simulations of traffic flow, $kT$ is set to be $1/\rho$, and $\rho$ represents the density. Substituting Eq. (\ref{Eq:Hamiltonian_Global}) into Eq. (\ref{Eq:Partition function 1}), we obtain a detailed expression of the partition function:
\begin{align} 
\begin{aligned}\label{sec:energy from field2}
    Z = \sum_{\mathbf{S}} \mbox{exp}\left\{\frac{J}{kT}\sum_{<i,j>}^N S_iS_j + \frac{F}{kT}\sum_{i=1}^N S_i\right\}
\end{aligned}
\end{align}

Before further simplifying the partition function, we may rewrite the Eq. (\ref{sec:energy from field2}) as:

\begin{align} 
\begin{aligned}\label{sec:energy from field}
    Z &= \sum_{\mathbf{S}} \mbox{exp}\left\{K\sum_{<i,j>}^N S_iS_j + B\sum_{i=1}^N S_i\right\}\\
\end{aligned}
\end{align}

\noindent where $K = J/kT$ and $B = F/kT$ are functions of the vehicular density in the ring road traffic system. Assuming that the system only exhibits nearest neighbor interactions, the partition function may be written as:
\begin{align}
    Z &= \sum_{\mathbf{S}} \mbox{exp}\left\{K\sum_{i}^N S_iS_{i+1} + B\sum_{i=1}^N S_i\right\}\nonumber \vspace{1em}\\
    \implies Z&= \sum_{\mathbf{S}} \prod_{i \in N}\mbox{exp}\:\{K S_iS_{i+1}+BS_i\}
     \label{Eq:total partition function}
\end{align}

Renormalization group transformations can be applied to this partition function to obtain the interactions and field coefficients $K'$ and $B'$ for the coarse-grained representations of the original coefficients $K$ and $B$, respectively. These coarse-grained coefficients can then be used to simulate the traffic flow dynamics described in Algorithm 1, as discussed in the next few subsections and Appendix \ref{Sec:Appendix}. 




\subsection{Coarse-graining and Free Energy Invariance}
We obtain coarse-grained representations of the partition function $Z$ to model spatial scales larger than the original 5 m long fine-grained sites \cite{wilson1975renormalization}.
Thus the coarse-grained sites can now accommodate more than one vehicle, and consequently the representation of the `microstates' of the system will also change from fine-grained site lattices to coarse-grained block lattices. However, the free energy of the fine-grained model is expected to be the same as the free energy of the coarse-grained counterpart, since they are both representations of the same ring road traffic system. The invariance of the total free energy of the traffic system is expressed as:
   \begin{equation}
      \mathcal{F}'(\lambda') = -ln Z' \simeq  -ln Z =  \mathcal{F}(\lambda)
    \end{equation}
where $\lambda$ represents the finer spatial scale, $\lambda'>\lambda $ represents a spatial scale that is coarser than the fine-grained spatial scale $\lambda$, $\mathcal{F}(\lambda)$ denotes the total free energy of the ring road traffic system at the fine-grained spatial scale $\lambda$,  and $Z$ and $Z'$ denote the partition functions for spatial scales $\lambda$ and $\lambda'$, respectively. We use the notation $\gamma = \lambda'/\lambda$ to denote the scaling factor in successive RG transformations.
    
The RG transformation decimates spin degrees of freedom while preserving the total free energy. At each step that the RG transformation is applied to the traffic flow `spin' system, the number of sites in the system decreases from $N \to N'$, where $N' = N/\gamma^r$ in $r$-dimensional space. In our work, the ring-road represents a one-dimensional traffic system with periodic boundary conditions. Consequently, the number of sites before and after the RG transformation are related as $N' = N/2$, where the scale changes by a factor of $\gamma = 2$.  
Each coarse site is, of course, subsequently re-scaled by a factor of 2 as well, so that the coarse-grained system is similar to the original fine-grained system \cite{wilson1975renormalization}. 
Following Suzuki \cite{suzuki1977static}, we also perform coarse-graining of time such that $t' = t/2$, which helps retain the dynamical nature of traffic flow in the renormalized model.

\subsection{Renormalization Group Applied to Statistical Mechanics-inspired Traffic Flow Model}
\label{Sec:RG-approach-to-SMIT}
The goal of the application of renormalization group transformations to the statistical mechanics-inspired traffic flow model is to obtain the coarse-grained interaction and field coefficients (i.e., $K'$ and $B'$, respectively) that govern traffic flow dynamics. These coefficients can then be used to model and simulate traffic flow at different spatiotemporal scales. Appendix \ref{Sec:Appendix} provides details on how to relate the fine-grained partition function $Z$ to the coarse-grained partition function $Z'$ with a similar functional form, and obtain the renormalized coefficients. Concisely, we re-label spins in the fine-grained partition function shown in Eq. (\ref{Eq:total partition function}) as $i = 1, 2, 3, . . . , N/2$. Next, a partial sum is performed over all the even-numbered sites, effectively decimating the odd-numbered  sites or cells on the road, to yield the following form of the partition function:
\begin{align}
    Z = \sum_{\{...S_1,S_3,S_5...\}} \prod_{i=2,4,6,...} & \sum_{\{S_i\}} \mbox{exp}\left\{K S_{i-1}S_{i}+ K S_iS_{i+1}+B\left(\frac{S_{i-1}}{2}+S_i+\frac{S_{i+1}}{2}\right)\right\}
\label{Eq:Z-fine-grained}
\end{align}
Since the fine and coarse-grained partition function represent the same system, the corresponding partition functions are also equivalent. The coarse-grained partition function may be written in the following form:
\begin{align} 
    Z' &= \sum_{\mathbf{S}} \prod_{i} f(K,B)\:\mbox{exp}\left\{K'S_{i-1} S_{i+1}+B'\left(\frac{S_{i-1}}{2}+\frac{S_{i+1}}{2}\right)\right\}
\label{Eq:Z-coarse-grained}
\end{align}
\noindent where $K$ represents the interaction coefficient between neighboring sites on the roadway, $B$ reflects the field coefficient biasing the forward motion of vehicles, $f(K,B)$ is a nonlinear function of parameters $K$ and $B$, and $\mathbf{S}$ represents all the possible state combinations for sites with odd index. 
As Appendix \ref{Sec:Appendix} shows, Eqs. (\ref{Eq:Z-fine-grained}) and (\ref{Eq:Z-coarse-grained}) may be used to obtain the coarse-grained coefficients $K'$ and $B'$ after the RG transformation, to yield the following results:
\begin{align}
    B' &= \frac{1}{2}ln \left\{ \frac{e^{\{2K+2B\}} + e^{\{-2K\}}}{e^{\{2K-2B\}}+e^{\{-2K\}}} \right\}{\label{Eq:Renormalized-coeffs-B}}\\
    K' &= \frac{1}{4}ln\frac{(e^{\{2K+2B\}} + e^{\{-2K\}})(e^{\{-2K\}}+e^{\{2K-2B\}})}{e^{\{-2B\}}+2+e^{\{2B\}}}
    {\label{Eq:Renormalized-coeffs-K}}
\end{align}


This approach provides an analytical solution to the renormalization group approach applied to traffic flow, which has also been pursued via numerical means by other researchers \cite{teoh2018renormalization}. Parameters for subsequent coarse-grained spatial scales (e.g. ($K'', B''$), ($K''', B'''$), and so on)  can be obtained using Eqs. (\ref{Eq:Renormalized-coeffs-B}) and (\ref{Eq:Renormalized-coeffs-K}) recursively.  The next section demonstrates the results from numerical simulations that use the analytically-derived model parameters $K'$ and $B'$ when multiple RG transformations are applied recursively to the CA-based traffic flow model.

\section{Results}
\label{Sec:Results}
A key insight of the presented work is that the \textit{coarse-grained} model parameters obtained via RG transformations produce the \textit{same} emergent patterns and dynamics of traffic flow, even at different spatiotemporal scales. Specifically, with each RG transformation, we obtain successively coarse-grained descriptions of the model as well as the model parameters $K'$ and $B'$ as shown in Eqs. (\ref{Eq:Renormalized-coeffs-B}) and (\ref{Eq:Renormalized-coeffs-K}). These coarse-grained parameters are substituted into the local Hamiltonian $H'_s$ to obtain transition probability of forward motion on the roadway: \begin{align}
    \phi = exp(-\beta H'_s)
\end{align}
The transition probability is then used to simulate the traffic flow system at coarser scales according to Algorithm 1. The results of these simulations are discussed below.

\subsection{Coarse-grained Traffic Simulations using Renormalization Group Transformations}
Traffic flow simulations were conducted on a ring road environment approximately 1.25 km in length, and at two different vehicular densities representative of free flow and congested traffic. The RG transformation were performed to obtain the coarse-grained model parameters for five different spatial scales, corresponding to spatial resolutions of 5 meters/site, 10 m/site, 20 m/site, 40 m/site, and 80 m/site. Thus, at successive coarse-graining levels, the descriptive length of a site in the CA model was increased by a factor of two, which also corresponds to halving the number of sites per RG transformation (since the length of the ring road environment remains unchanged). Moreover, the initial conditions of the system were also coarse-grained through decimation to ensure similarity across spatial scales, both in terms of density and initial system configuration. The coarse-grained model parameters and initial conditions were then used to simulate the evolution of the coarser system. The resulting traffic flow dynamics were compared against the original simulation.

\begin{figure}[!ht]
\centering
  \includegraphics[width=0.7\linewidth]{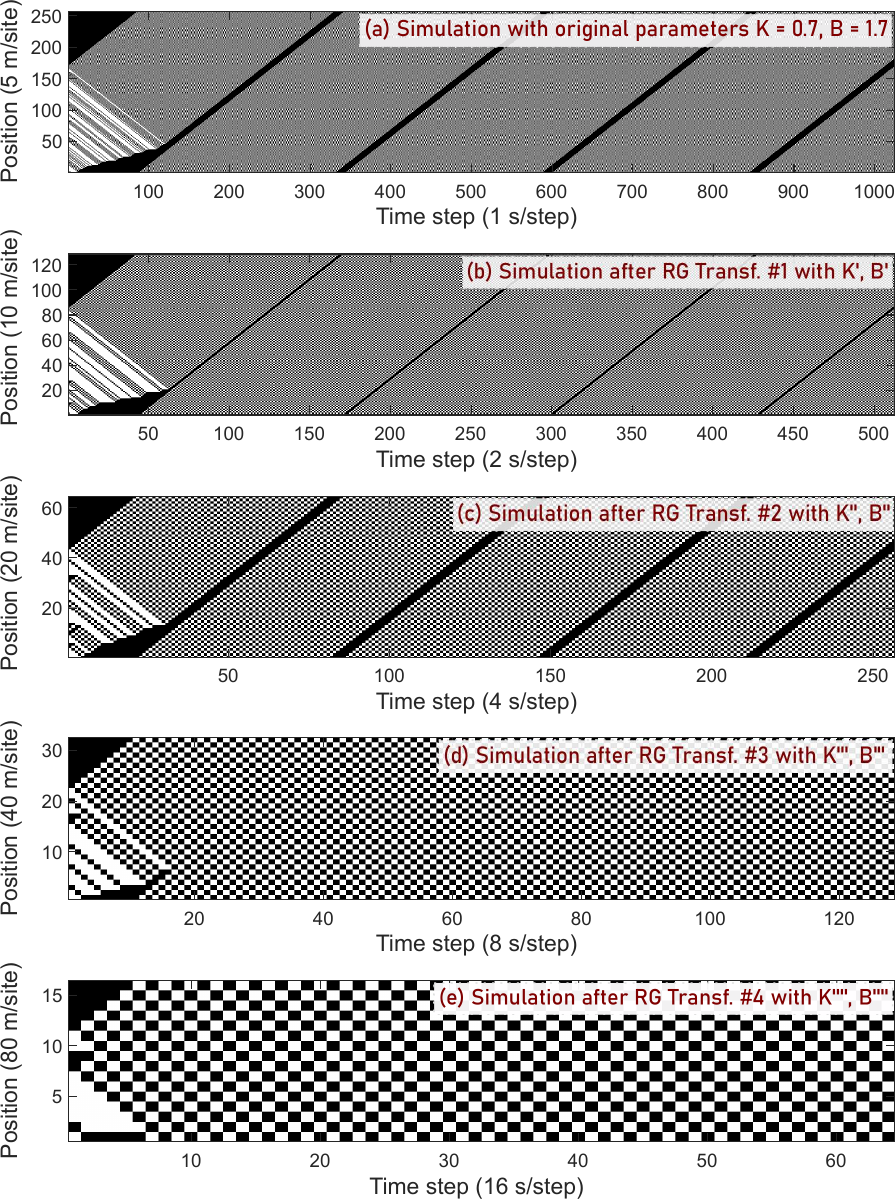}
	\caption{Traffic flow simulations representing \textbf{free flow at medium density} ($\rho \approx 0.5$) with the original model parameters (i.e. interaction and field coefficients) of $K = 0.7$ and $B = 1.7$, respectively. Forward motion of vehicles is visible. Simulations are successively coarse-grained by a factor of 2 via Renormalization Group transformations. Sites at successive coarse-grained scales span 5 m/site (original scale), 10 m/site, 20 m/site, 40 m/site, and 80 m/site. Note that both spatial and temporal scales are coarse-grained in the simulations.}
	\label{fig:Traffic flow with samller Cof}
\end{figure}

\begin{figure}[!ht]
\centering
  \includegraphics[width=0.7\linewidth]{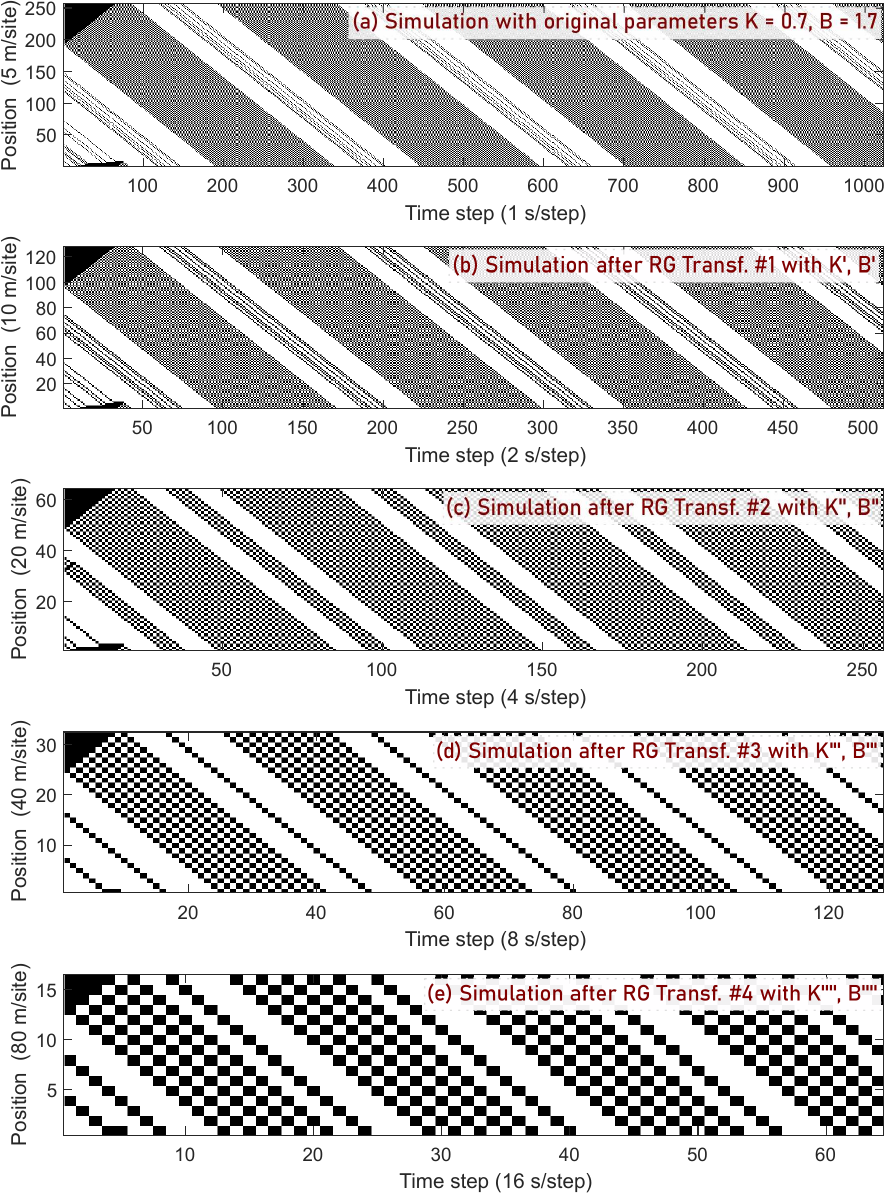}
	\caption{Traffic flow simulations representing \textbf{congested flow at high density} ($\rho \approx 0.7$) with the original model parameters (i.e. interaction and field coefficients) of $K = 0.7$ and $B = 1.7$, respectively. Backward moving congestion waves (white) are visible. Simulations are successively coarse-grained by a factor of 2 via Renormalization Group transformations. Sites at successive coarse-grained scales span 5 m/site (original scale), 10 m/site, 20 m/site, 40 m/site, and 80 m/site. Note that both spatial and temporal scales are coarse-grained in the simulations.}. 
	\label{fig:Traffic flow with larger Cof}
\end{figure}

Figs. \ref{fig:Traffic flow with samller Cof} and \ref{fig:Traffic flow with larger Cof} show the simulated traffic flow dynamics for different densities within the same initial parameter set, i.e. $K = 0.7$ and $B = 1.7$. Fig. \ref{fig:Traffic flow with samller Cof} corresponds to free flow traffic dynamics observed in medium density traffic ($\rho \approx 0.5 $), and is also representative of dynamics at densities lower than 0.5. It demonstrates transient phenomena, such as the disappearance of an initially congested or queued state by about 100 seconds into the simulation. The five plots in Fig. \ref{fig:Traffic flow with samller Cof} represent simulations performed with  coarse-grained parameters, i.e. $K'$ and $B'$ obtained after the first RG transformation, $K''$ and $B''$ obtained after the second RG transformation, and so on. Similar traffic dynamics and trends are observed in all plots, such as the similar duration for which the initially congested state persists. Additional discussion on the similarity in traffic system evolution across different scales is included in Section \ref{sec:Accurcacy}.
Fig. \ref{fig:Traffic flow with larger Cof} demonstrates the simulated traffic flow dynamics for density $\rho = 0.7$. At densities higher than the critical density, traffic jams appear consistently and persist.
 As expected, results in Fig. \ref{fig:Traffic flow with larger Cof} show the emergence of backward moving congestion waves \cite{Daganzo_CA_KWT}. Additionally, patterns observed in simulations at the original spatial resolution are retained at coarser spatial scales which utilize renormalized parameters. 

 The parameters $K'$ and $B'$ obtained analytically after each RG transformation can be used to simulate the dynamics of traffic flow at successively coarser spatiotemporal scales. This approach will enable practitioners and researchers to relate microscopic and macroscopic models of traffic flow at several different spatial scales, including at arbitrarily chosen spatial scales \cite{Yang2020}. More importantly, free energy invariance inherent in this approach is expected to preserve the emergent traffic flow patterns across spatial scales \cite{friston2010free}. This enables macroscopic-scale traffic predictions using only coarse-grained simulations, even though these contain less information that the fine-grained simulations.
 The next subsection discusses the accuracy of the coarse-grained simulations as compared to simulations performed at the original spatial resolution.


\subsection{Model Accuracy after Renormalization Group Transformations}
\label{sec:Accurcacy}

\begin{figure}[!ht]
\centering
  \includegraphics[width=0.7\linewidth]{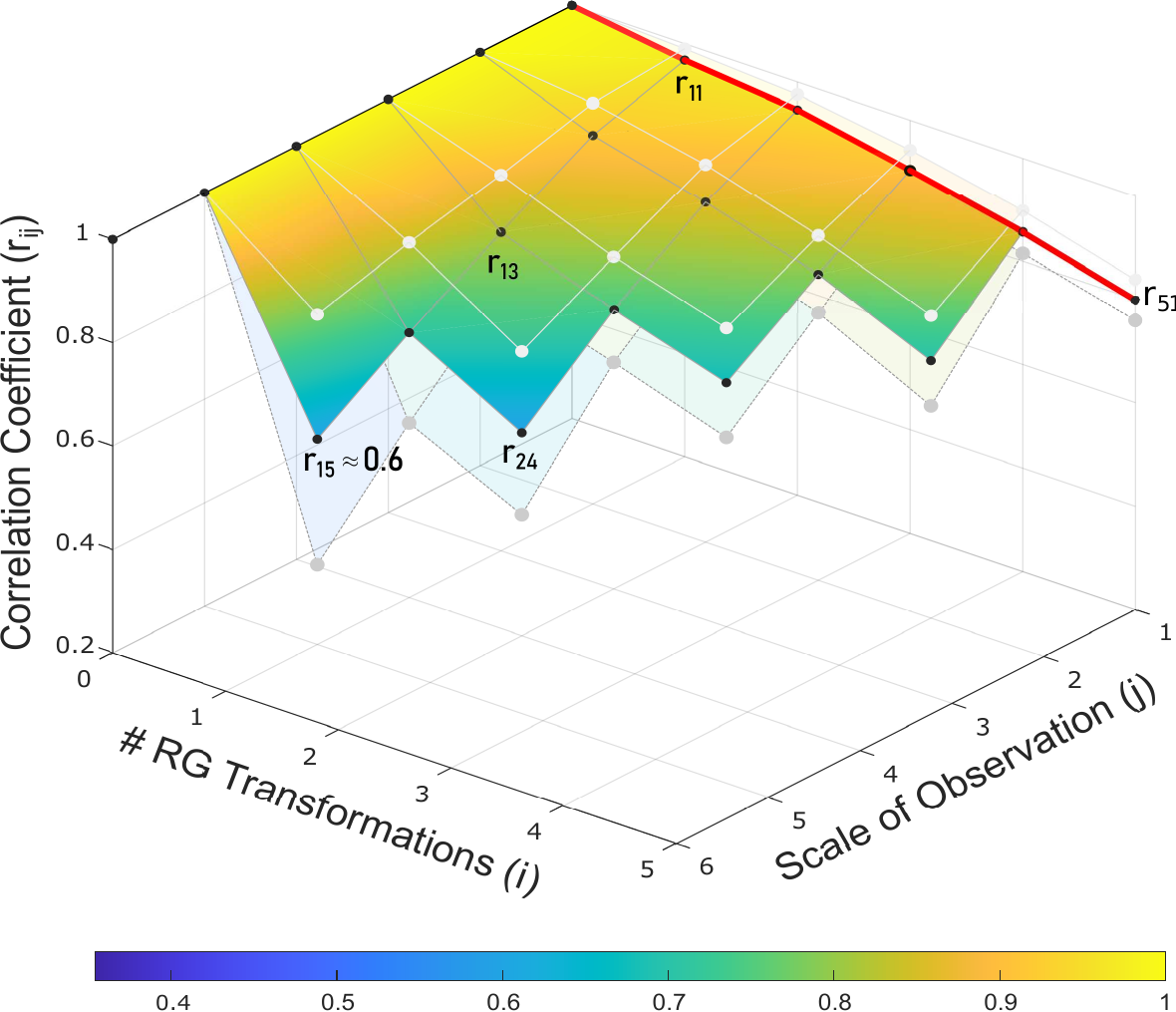}
	\caption{Correlation coefficients indicating accuracy of coarse-grained traffic simulations, obtained by pixel-based comparison of finer-grained and coarser-grained `images'. Mean correlation coefficients (black dots) show a decreasing accuracy with increasing number of RG transformations, but a reasonable accuracy (thick red line; $r_{i1}$) is retained even after several RG transformations on the finest-grained simulation ($j = 1$). One standard deviation values of correlation coefficients is shown with grey dots.}
	\label{Fig: Accuracy vs Information at different scales}
\end{figure}

With each RG transformation, some information about the traffic microstate is lost. Thus, if coarse-grained models are to be used for traffic flow prediction or control, it is pertinent to know how accurately the coarse-grained simulations represents the actual traffic flow \cite{giuliani2021gentle}.
Our approach towards evaluating accuracy across multiple spatial scales relies on measuring correlation between pixel-based matrices, i.e. the time-space `images' at different stages of the RG transformations as shown in Figs. \ref{fig:Traffic flow with samller Cof} or \ref{fig:Traffic flow with larger Cof}.
Fig. \ref{Fig: Accuracy vs Information at different scales} represents the results from 100 traffic flow simulations performed at each combination of scale of observation ($i$) and RG transformations ($j$). The black dots represent the mean correlation coefficient from these simulations, with a color-mapped surface added to aid in interpretation. The light grey dots represent one standard deviation of the correlation coefficient from the mean, as observed in the simulations.
The correlation is quantified using the following expression for the correlation coefficient:

\begin{equation}
    r_{ij} = \frac{\sum\limits_m \sum\limits_n(p^{mn}_j-\bar{p}_j)(p^{mn}_i-\bar{p}_i)}{\sqrt{\big(\sum\limits_m \sum\limits_n (p^{mn}_j-\bar{p}_j)^2\big)\big(\sum\limits_m \sum\limits_n(p^{mn}_i-\bar{p}_i)^2\big)}}
    \label{Eq:Correlation}
\end{equation}  
where  $p^{mn}_j \in \{0,1\}$ represents each pixel value in time-space diagram of a fine-grained system (i.e. at the $j^{th}$ level), $\bar{p}_j\in\mathbb{R}$ represents the mean pixel value in the image, $m$ represents the index corresponding to time, $n$ represents the index for space, $p^{mn}_i$ represents each pixel value in time-space diagram of a coarse-grained system (i.e. after $i$ RG transformations), and $r_{ij}$ denotes the image correlation coefficient between traffic simulation images obtained before and after the $i^{th}$ RG transformation applied to the $j^{th}$ spatial scale. For example, $r_{11}$ represents the correlation coefficient between simulation images at the original resolution or scale of observation ($j=1$) and those obtained after one RG transformation ($i=1$) of the original system. Similarly, $r_{24}$ represents the correlation coefficient between traffic simulation images at the fourth coarse-graining level ($j=4$) and those obtained after two RG transformations ($i=2$) on the fourth coarse-graining level.
Moreover, since the number of sites and time steps is halved after each RG transformation, it is not possible to directly compare images at two different coarser-grained levels. For example, Fig. \ref{fig:Traffic flow with larger Cof}(a) represents the simulation image at the original resolution ($j=1$) and has $1024 \times 256$ pixels, whereas Fig. \ref{fig:Traffic flow with larger Cof}(b) obtained after one RG transformation ($i=1$) has an image resolution of $512 \times 128$ pixels. 
To enable a pixel-by-pixel comparison and correlation analysis, the coarse-grained simulation images were resized to match the size of the finer-grained simulation images. Thus, $p^{mn}_i$ in Eq. (\ref{Eq:Correlation}) represents the modified pixel-based matrix corresponding to the resized image of the coarser-grained traffic flow simulation.

Clearly, the correlation coefficients ($r_{ij}$) depend on the scale at which simulations are performed or observed ($i$), as well as number of RG transformations applied ($j$) to the dynamics at this scale. Fig. \ref{Fig: Accuracy vs Information at different scales} describes how correlation coefficients are influenced by these two variables. In this figure, the scale of observation represents the resolution of a traffic simulation (coarser or finer), against which subsequent images of further coarser-grained traffic flow dynamics (obtained via RG transformations) are compared. For example, Fig. \ref{fig:Traffic flow with larger Cof}(c) may be the original scale of observation ($j=3$), i.e. the simulation is performed using a resolution of 20 m/site and 4 s/step, which corresponds to traffic dynamics generated using model parameters $K''$ and $B''$. Applying one RG transformation ($i=1$) at this level, will yield a coarser traffic dynamics simulation at coarse-graining level (Fig. \ref{fig:Traffic flow with larger Cof}(d)), and comparing these two simulation images will provide the correlation coefficient $r_{13}$, as indicated in Fig. \ref{Fig: Accuracy vs Information at different scales}. 


These correlation coefficients ($r_{0j}$) effectively indicate correlation of an image with itself, and are always equal to 1.
The mean correlation coefficients $r_{i1}$ represent the comparison of the various coarser-grained traffic simulation images obtained via recursive application of RG transformations against the original fine-grained simulation (performed with a spatial resolution of 5 m/site and 1 s/step). These coefficients $r_{i1}$ are indicated via the thick red line in the figure. The simulation results indicate that traffic simulation performed at coarser scales (using model parameters obtained after RG transformations) still maintain a good level of accuracy when compared against the finest-grained traffic simulations. In fact, even after five RG transformations on the finest-grained system, the mean correlation coefficient $r_{51}$ is approximately 0.8, with a small standard deviation. The lowest mean correlation coefficient across all simulations was found to be $r_{15} \approx 0.6$.

\subsection{Computational Complexity of Coarse-grained Simulations}
\begin{figure}[h]
\centering
  \includegraphics[width=0.6\linewidth]{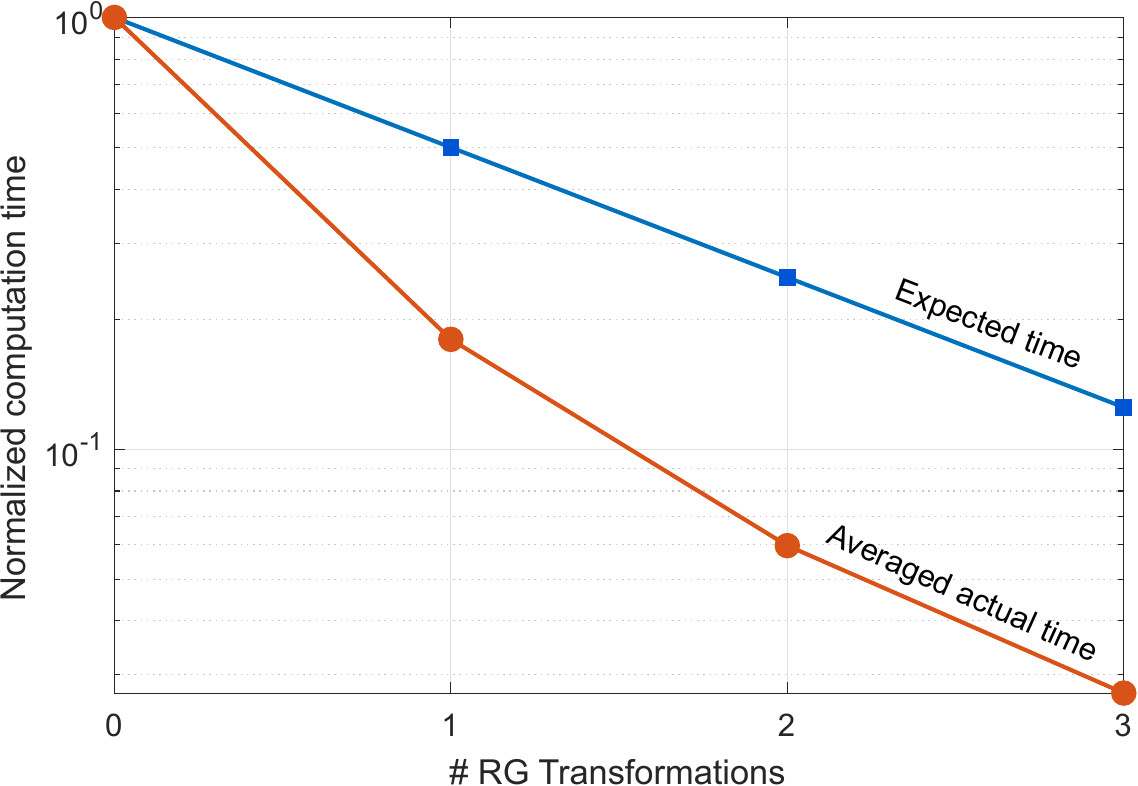}
	\caption{Expected and Averaged actual computation time, both normalized and represented in log scale. Both show decreasing trends as more RG transformations are applied, with the actual computation time decreasing faster than logarithmically.}
	\label{fig:Time complexity}
\end{figure}

The ability to simulate traffic dynamics at coarser spatiotemporal scales opens avenues for speeding up computation to enable real-time traffic flow prediction and control. 
Since the primary computation is associated with nearest neighbor interactions, the computational complexity for the finest-grained simulation is given by $\mathcal{O}(N\times T)$, where we use the Big-$\mathcal{O}$ notation.
The coarse-grained system obtained after the RG transformation has fewer sites and time steps, both reduced by the scaling factor $\gamma$. Thus, the computation complexity of the renormalized system is reduced by a factor of $\gamma^2$, and is given by:
\begin{equation}
  \mathcal{O}(N/\gamma \times T/\lambda) = \frac{1}{\gamma ^2} \mathcal{O}(N\times T)  
\end{equation}
As discussed before, if two sites are decimated to one after each RG transformation, the scaling factor is $\gamma =2$.
Fig. \ref{fig:Time complexity} shows the actual and expected computation times on a logarithmic scale for $\gamma=2$, averaged over the set of simulations discussed earlier. The simulation times are normalized so that the computation time associated with the fine-grained simulation is 1. It is observed that the time taken to simulate coarse-grained traffic systems will decrease faster than the log function. 

\section{Concluding Remarks} 
\label{Sec:4-Concluding-remarks}

Prior modeling approaches to traffic flow have demonstrated that there is value to modeling traffic flow at different scales, from microscopic to macroscopic. However, 
the research community is still seeking a solution to link together modeling at different spatial scales under a common systematic framework. Our work demonstrates that this is possible using the renormalization group theory and a statistical mechanics-inspired approach, enabling us to model traffic flow at a spatiotemporal scale of our choosing. More importantly, this can be done without changing the underlying modeling and simulation framework, while preserving accuracy. 
Moreover, we have shown that the emergent dynamic patterns such as traffic jams are reproducible at different spatiotemporal scales using this modeling approach. Further, based on simulation and modeling requirements, the spatiotemporal scales can be changed by a trade-off between  decreasing computation time or increasing accuracy. The presented work also demonstrates that the computation time associated with completing coarse-grained simulations decreases at a faster-than-logarithmic rate.

The results of this paper could enable faster, improved, and scale-dependent simulations of traffic flow dynamics, with potential benefits to practitioners, researchers, and engineers working with large-scale traffic systems. For future work, the authors intend to leverage this modeling approach to evaluate observability measures associated with making measurements at different spatial scales, as well as using coarse-grained models to make traffic flow predictions, as well as examine singular behaviors of this approach.

\section*{Acknowledgements}
This material is based upon work supported by the National Science Foundation under Grant No. 1663652. Any opinions, findings, and conclusions or recommendations expressed in this material are those of the authors and do not necessarily reflect the views of the National Science Foundation.

\begin{appendices}
\section{Renormalization Group on Statistical Mechanics-based Traffic Flow }
\label{Sec:Appendix}
Here we show how the recursive relationship between interaction and field coefficients is derived. In other words, we demonstrate how to obtain the coarse-grained model parameters $K'$ and $B'$ as functions of the finer-grained model parameters $K$ and $B$. Section \ref{Sec:RG-approach-to-SMIT} already established that since the fine-grained and coarse-grained models are representations of the same system, their free energy and partition functions should be invariant, such that $Z = Z'$. The expressions for $Z$ and $Z'$ are reproduced here for convenience:
\begin{align}
  \begin{aligned}\label{GR on total energy}
    Z = \sum_{\{...S_1,S_3,S_5...\}} \prod_{i=2,4,6,...} & \sum_{\{S_i\}} \mbox{exp}\left\{K S_{i-1}S_{i}+  K S_iS_{i+1}+B\left(\frac{S_{i-1}}{2}+S_i+\frac{S_{i+1}}{2}\right)\right\}
\end{aligned}  
\end{align}
\noindent where the spins have been relabeled as $i = 1, 2, 3, ... N/2$, and a partial sum is performed over all even-numbered sites (thus decimating the odd-numbered sites). Additionally, we require the coarse-grained partition function $Z'$ to take the form:
\begin{align} 
\begin{aligned}\label{simplified total energy}
    Z' &= \sum_{\{\mathbf{S}\}} \prod_{i} f(K,B)\mbox{exp}\left\{K'S_{i-1} S_{i+1}+B'\left(\frac{S_{i-1}}{2}+\frac{S_{i+1}}{2}\right)\right\}
\end{aligned}
\end{align}
Equating these two forms of the partition function in Eq. (\ref{GR on total energy}) and Eq. (\ref{simplified total energy}), i.e. setting $Z = Z'$, yields:
\begin{align}
\label{Eq:Appendix-Partition-fn-comparison}
\begin{aligned}
   & \sum_{\{S_i\}}\mbox{exp}\left\{K(S_{i-1}S_i+S_i S_{i+1})+B\left(\frac{S_{i-1}}{2}+S_i+\frac{S_{i+1}}{2}\right)\right\} \\ = &
   f(K,B)\mbox{exp}\left\{K'S_{i-1} S_{i+1}+B'\left(\frac{S_{i-1}}{2}+\frac{S_{i+1}}{2}\right)\right\}  
 \end{aligned}
\end{align}
We can further expand the expression on the left-hand side of Eq. \ref{simplified total energy} by incorporating the fact that $S_i$ can only take values in the set $\{-1, +1\}$, to yield:
\begin{align}\label{Eq:Appendix-partition-fn-original}
\begin{aligned}
   & \sum_{\{S_i\}}\mbox{exp}\left\{K(S_{i-1}S_i+S_i S_{i+1})+B\left(\frac{S_{i-1}}{2}+S_i+\frac{S_{i+1}}{2}\right)\right\} \\ = &  \mbox{exp}\left\{K(S_{i-1}+S_{i+1})+B\left(\frac{S_{i-1}}{2}+1+\frac{S_{i+1}}{2}\right)\right\} + \\
   &  \mbox{exp}\left\{-K(S_{i-1} +S_{i+1})+ B\left(\frac{S_{i-1}}{2}-1+\frac{S_{i+1}}{2}\right)\right\}
  \end{aligned}
\end{align}

Now, the right-hand sides of both Eqs. (\ref{Eq:Appendix-Partition-fn-comparison}) and (\ref{Eq:Appendix-partition-fn-original}) represent the same quantities \textit{and} are also expressed using the same variables ($S_{i-1}$ and $S_{i+1}$). We can now consider the explicit cases for the various possible configurations of $S_{i-1}$ and $S_{i+1}$. Specifically, in the context of Eqs. (\ref{Eq:Appendix-Partition-fn-comparison}) and (\ref{Eq:Appendix-partition-fn-original}), three distinct cases arise:

\noindent \underline{\textbf{Case 1}}: In this case, $S_{i-1} = S_{i+1} = +1$, i.e. both sites are in state $+1$. In this case, comparing the right-hand sides of Eqs. (\ref{Eq:Appendix-Partition-fn-comparison}) and (\ref{Eq:Appendix-partition-fn-original}) yields:
\begin{align}\label{sec:case 1}
\begin{aligned}
    \mbox{exp}\{2K+2B\} + \mbox{exp}\{-2K\} =  f(K,B)\mbox{exp}\{K'+B'\}
  \end{aligned}
\end{align}
\noindent \underline{\textbf{Case 2}}: Here, we assume that $S_{i-1} = S_{i+1} = -1$, i.e. both are in state $-1$. Again, performing a similar comparison as in Case 1, we get:
\begin{align} \label{sec:case 2}
\begin{aligned}
    \mbox{exp}\{-2K\} + \mbox{exp}\{2K-2B\} =  f(K,B)\mbox{exp}\{K'-B'\}
  \end{aligned}
\end{align}
\noindent \underline{\textbf{Case 3}}: The final scenario is where $S_{i-1} = -S_{i+1} = \pm 1$, i.e. the two sites have states that are different from the other. This yields:
\begin{align}\label{sec:case 3}
\begin{aligned}
    \mbox{exp}\{B\}+\mbox{exp}\{-B\} =  f(K,B)\mbox{exp}\{-K'\}
  \end{aligned}
\end{align}

Now, we can perform arithmetic operations on equations obtained from each of these three cases to obtain the expression for $f(K,B)$, as well as the two recursion relationships that relate $K'$ and $B'$ to the coefficients $K$ and $B$. Specifically, dividing Eq. (\ref{sec:case 1}) by Eq. (\ref{sec:case 2}) and simplifying, we can derive an expression for $B'$ as follows: \begin{align} {\label{B_prime}}
\begin{aligned}
    B' = \frac{1}{2}ln \left\{ \frac{e^{\{2K+2B\}} + e^{\{-2K\}}}{e^{\{2K-2B\}}+e^{\{-2K\}}} \right\}
  \end{aligned}
\end{align}
Similarly, dividing the product of Eqs. (\ref{sec:case 1}) and (\ref{sec:case 2}) by the square of Eq. (\ref{sec:case 3}) and then simplifying, the expression for $K'$ can be obtained as follows: 
\begin{align}
\begin{aligned}
    K' = \frac{1}{4}ln\frac{(e^{\{2K+2B\}} + e^{\{-2K\}})(e^{\{-2K\}}+e^{\{2K-2B\}})}{e^{\{-2B\}}+2+e^{\{2B\}}}
  \end{aligned}
\end{align}
Finally, the function $f(K,B)$ can be calculated from the renormalized parameters in a relatively straightforward fashion:
\begin{align}
\begin{aligned}
    f(K,B) = exp\{K'+B\}+exp\{K'-B\}
  \end{aligned}
\end{align}
where $K'$ can be substituted into the above equation to obtain an expression for $f(K, B)$ in terms of $K$ and $B$ alone. The model parameters $K'$ and $B'$ now enable us to perform coarse-grained traffic flow simulations at different spatiotemporal scales as discussed in Section \ref{Sec:Results}.
\end{appendices}

\bibliographystyle{unsrt}  

\bibliography{references}

\end{document}